\documentclass{icrc2009}

\usepackage{graphicx}   
\usepackage{caption}    
\usepackage[font=footnotesize]{subfig} 
\usepackage{fixltx2e}
\usepackage{multirow}
\usepackage{url}

\newcommand{\shorttitle}[1]%
{\markboth{Proceedings of the 31\MakeLowercase{$^{st}$} ICRC, {\L}\'{o}d\'{z} 2009}{#1} }
\newcommand{\etal}{\MakeLowercase{\textit{et al. }}} 

\begin{document}
\title{The VERITAS Survey of the Cygnus Region of the Galactic Plane}

\author{\IEEEauthorblockN{Amanda Weinstein\IEEEauthorrefmark{1},
              for the VERITAS Collaboration\IEEEauthorrefmark{2}}
              \\
\IEEEauthorblockA{\IEEEauthorrefmark{1} University of California Los Angeles,
475 Portola Plaza, Los Angeles, CA 90095 USA}
\IEEEauthorblockA{\IEEEauthorrefmark{2} see R. A. Ong et al. (these
proceedings) or http://veritas.sao.arizona.edu/conferences/authors?icrc2009}}

\shorttitle{Weinstein \etal VERITAS Survey} \maketitle

\shorttitle{Weinstein \etal VERITAS Survey} \maketitle

\begin{abstract}
 The Cygnus region of the Galactic plane contains many known supernova remnants, pulsars, X-ray and GeV
 gamma-ray emitters which make it a prime candidate for a Very High Energy (VHE) gamma-ray
 survey in the Northern Hemisphere.
 The VERITAS observatory, an array of four atmospheric Cherenkov telescopes located at the base of Mt. Hopkins
 in southern Arizona, USA, has carried out an extensive survey of the Cygnus region between 67 and 82 degrees in galactic
 longitude and between -1 and 4 degrees in galactic latitude. The survey, comprising more than 140 hours of observations,
 reaches an average VHE flux sensitivity of better than $\textbf{4\%}$ of the Crab Nebula at energies above 200 GeV.
 Here we report on the preliminary results from this survey.
  \end{abstract}


\section{Introduction}

To date, only a few moderate-scale surveys have been performed in gamma rays
between $\rm 100 \thinspace GeV$ and $\rm 10 \thinspace TeV$: the HESS scan of
the central region of the Galactic Plane\cite{hess_survey}, the much less
sensitive HEGRA survey of the quarter of the Galactic Plane between
$-2^{\circ}$ and $85^{\circ}$ in galactic longitude\cite{hegra}, and the
VERITAS scan of the Cygnus region under discussion here. However, there is a
strong motivation for performing such surveys; an unbiased search of a
substantial region of sky is less subject to the experimental and theoretical
prejudices that guide most VHE gamma-ray observations, and therefore (as
demonstrated by HESS) offers greater scope for serendipitous discoveries.  An
unbiased search also allows for more quantitative statements to be made about
the source population in the region surveyed.

The stereoscopic imaging atmospheric Cherenkov telescope (IACT) array VERITAS
has just completed a two-year survey of a 5 by 15 degree portion of the Cygnus
region of the Galactic Plane.  The Cygnus region was a natural target for
survey observations, as it is already known to contain a significant number of
potential TeV gamma-ray emitters.  In the GeV ($\rm 20 MeV-300$ GeV) energy
band, it is home to a number of sources or potential sources, including no less
than 4 distinct Fermi sources \cite{fermi}.  Moreover, both Fermi and its
predecessor EGRET have detected diffuse emission from this region that is
greater in flux than all of the currently resolved sources taken together.
Viewed in the energy range between 1TeV-50 TeV, it contains a pair of
unidentified TeV sources (MGRO J2031+41 and MGRO 2019+37) detected by the
Milagro Gamma Ray Observatory, a water Cherenkov extensive air shower array
that is sensitive to TeV sources at a median energy of 20 TeV over the entire
sky \cite{Abdo:2007ad}, as well as the unidentified source TeV J2032+4130
(first detected by the HEGRA IACT array\cite{tev2032}), that is spatially
coincident with MGRO J2031+41.  The exact nature of these sources is currently
unknown. There is also a significant catalog of objects detected at other
wavelengths, including including SNRs, pulsar wind nebulae (PWNe), high-mass
x-ray binaries (HMXBs) and massive star clusters, that are considered potential
TeV sources.

\section{Survey Observations}

\begin{figure*}[!t]
  \centering
  \includegraphics[width=5in]{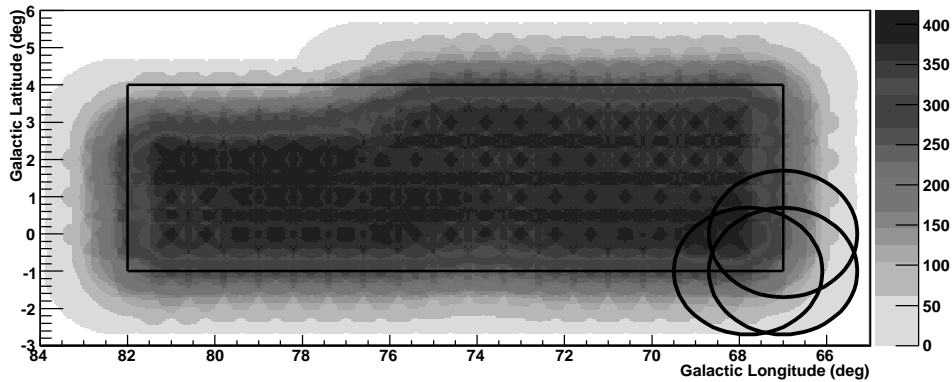}
  \caption{Effective exposure map for the VERITAS Cygnus sky survey,
based on data obtained from Spring 2007 through Fall 2008. The color scale to
the right gives the effective exposure time in minutes.  The black box
indicates the boundary of the survey region proper; the black circles show
examples of the overlapping fields of observation that are tiled to produce the
survey.}
  \label{survey_exposure}
 \end{figure*}

Survey observations, which began in April 2007 and were completed in November
2008, cover the field between galactic longitudes $67\,^{\circ}$ and
$82\,^{\circ}$ and galactic latitudes $-1\,^{\circ}$ and $4\,^{\circ}$ with a
grid of pointed observations. Grid points have $0.8\,^{\circ}$ separation in
Galactic latitude and $1\,^{\circ}$ separation in galactic longitude, allowing
for substantial overlap in the fields of view for observations at nearby grid
points.  Approximately one hour of observing time is taken at every grid point
(generally within a 1-3 day period), with that hour broken into 20-minute
observation periods. Figure \ref{survey_exposure} shows both a schematic of the
survey observation strategy and the acceptance-corrected (``effective'')
exposure time over the entire survey field. The base survey achieves a
relatively uniform effective exposure of $\sim 6$ hours.  The full survey
dataset ($>140$ hours of good-quality observation time) has regions of enhanced
exposure due to follow-up observations (some taken in fall 2008, others
scheduled to be taken in spring 2009).

Survey data have been quality-selected to remove runs with unstable trigger
rates, poor weather conditions, and known hardware problems. Almost all survey
data were taken on moonless nights, with a few percent taken under slight to
moderate moonlight conditions. Since survey observations began during the
commissioning period for the VERITAS array, data in the survey proper were
taken in two different configurations. Observations from spring 2007 were taken
with a three-telescope array configuration (including telescopes 1, 2, and 3)
and observations from fall 2007 and spring and fall 2008 were taken with the
full four-telescope array.  Follow-up data taken in fall 2008 were also taken
with the four-telescope array configuration; due to a scheduled VERITAS
upgrade, any follow-up data taken in Spring 2009 will be taken with the
three-telescope array configuration containing telescopes 2, 3, and 4.

\subsection{Zenith Angle of Observations}

Survey observations were taken over a range of zenith angles between
$10^{\circ}$ to $35^{\circ}.$  Observations were scheduled in such a way as to
keep the average zenith angle of observations for any survey pointing at
$20^{\circ}$, but constraints in terms of available time did not always allow
for this.  We estimate that $45\%$ of the survey region is covered by
observations taken at an average zenith angle of $20^{\circ}$ or smaller,
another $33.5\%$ by observations taken at $20^{\circ}  < z < 25^{\circ}$, $9\%$
by observations taken at $25^{\circ} < z <30^{\circ}$, and $12.5\%$ by
observations taken at $z > 30^{\circ}$.

\section{Survey Analysis}

Given that the H.E.S.S. survey revealed a population of Galactic TeV gamma-ray
emitters that was biased towards hard-spectrum, moderately extended
sources\cite{hess_survey}, and the fact that that the Milagro survey of the
Cygnus region \cite{Abdo:2007ad} showed sources of significant apparent extent,
it was reasonable to expect that some or all of the visible gamma-ray sources
in this region would also have relatively hard spectra and significant
extension.

In order to optimize the survey's sensitivity to such sources while not
sacrificing too much in the way of sensitivity to softer-spectrum and/or
point-like sources, a set of parallel analyses was used.  Each analysis is a
variation on a common base analysis, optimized for better sensitivity to a
particular type of source. In order to limit the number of additional trials
factors incurred, the number of parallel analyses was restricted to four.  Two
were optimized for sources with a Crab-like spectrum---one for point sources
and the other for moderately extended ($r=0.2^{\circ}$) sources---and two more
were optimized for harder spectrum point-like and extended sources, using a
reference source with a power-law spectrum and a spectral index of $2.0$.

The completely common elements of the analysis procedure consist of
calibrating and cleaning the Cherenkov images and parameterizing
them by second moments\cite{hillas}. The technique used to
stereoscopically reconstruct the shower direction and impact
parameter is likewise common to all analyses: however, images used
in stereoscopic reconstruction are required to exceed a minimum
integrated charge (\emph{size}) in digital counts (dc), and the
value of that requirement is analysis-specific as shown in Table
\ref{cut_summary}.

 \begin{table}[!h]
  \caption{Cuts Specific to Parallel Analyses}
  \label{cut_summary}
  \centering
  \begin{tabular}{|c|c|c|}
  \hline
   \hfil  &  Point Source & Extended Source \\ \hline
   \multirow{2}{*}{Soft Source}
    & $\rm size > 600 dc (\sim 90 p.e.)$  & $\rm size > 600 dc$ \\
    & $\theta^2 < 0.013$ &   $\theta^2 < 0.055$
    \\ \hline
   \multirow{2}{*}{Hard Source}
    & $\rm size > 1000 dc (\sim 150 p.e.)$  & $\rm size > 1000 dc$ \\
    & $\theta^2 < 0.013$ &   $\theta^2 < 0.055$
    \\ \hline

  \end{tabular}
  \end{table}

Cosmic ray background is rejected using two means.  The first is a pair of
variables (\emph{mean scaled length} (MSL) and \emph{mean scaled width} (MSW)
that summarize differences in image shape between gamma ray events and the
majority of the cosmic ray background)\cite{mslw} and are applied prior to
generation of photon sky maps.  The second is a cut on the square of the
angular distance ($\theta^2$) between a reconstructed shower and the sky
position of a potential source that is applied as part of that process.  The
cuts on \emph{mean scaled length} and \emph{mean scaled width} are common to
all parallel analyses ($\rm 0.05 < MSW < 1.06$, $\rm 0.05 < MSL < 1.24$), while
the cut in $\theta^2$ is analysis-specific as shown in Table \ref{cut_summary}.
The residual cosmic ray background is estimated using the ``ring-background''
model~\cite{ring}.

\section{Assessment of Survey Sensitivity}

\begin{figure}[!t]
  \centering
  \includegraphics[width=2.5in]{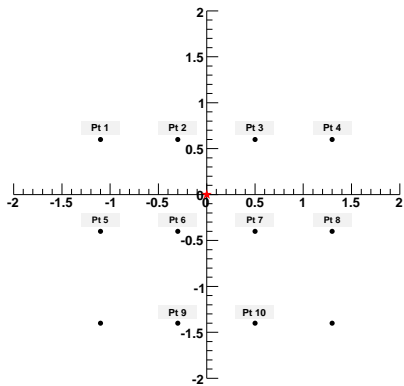}
  \caption{Structure of the simulated survey grid, with the contributing observation positions numerically
  labelled.  The test point is positioned at the origin.}
  \label{fig:cell}
 \end{figure}

A set of detailed survey simulations, coupled with observations of
the Crab Nebula at multiple offsets, allows us to determine the
\emph{a priori} sensitivity of the survey analyses, not only to a
point source with a Crab-like spectrum, but to harder-spectrum
and/or significantly extended sources.  In order to best reproduce
the expected background conditions, blank (i.e. cosmic-ray
dominated) survey fields were used to provide the background for
these simulations.  Blank fields at an appropriate range of zenith
angles were pulled from the survey and arranged in a mocked-up
`cell' of the survey grid around a test point, as shown in Figure
~\ref{fig:cell}. Showers that reconstruct at a distance greater than
$1.7^{\circ}$ from the center of the field of view are not used in
the analysis of survey data. Therefore, only pointings where the
center of the field of view is less than $1.7^{\circ}$ from the test
position are included in the simulation, as only these pointings
contribute significantly to the sensitivity at the test point.

Simulated gamma rays were then injected into each field in the simulated survey
grid at the test position at a rate appropriate to the source spectrum and flux
strength in question.  In each case the gamma rays were simulated at an
appropriate camera offset and matched as closely as possible to the background
field in terms of zenith angle and azimuth.  In the case of an extended source,
the injection positions were smeared by a two-dimensional Gaussian with
appropriate radii. The simulated grid is then analyzed using standard survey
analysis procedures.

In order to validate this technique, simulated Crab wobble observations, again
using survey background fields, were tested for consistency against standard
wobble ($0.5^{\circ}$ offset) and larger-offset Crab data; the predicted rates
and sensitivities appear to be in good agreement.

Preliminary results of the sensitivity studies, using a simulated survey grid
of $20^{\circ}$ zenith angle four-telescope observations, have been completed.
These suggest that the soft-spectrum point-source survey analysis is sensitive
at $5\sigma$ or better pre-trials to a source with a spectral index of 2.5 and
a flux $3\%$ that of the Crab Nebula; the hard-spectrum point-source survey
analysis is likewise sensitive to sources with a spectral index of 2.0 and an
integral flux of better than $3\%$ of the Crab Nebula above 200 GeV.  Early
studies of simulated hard-spectrum extended sources (Gaussian radius of
$0.2^{\circ}$) suggest that the survey analysis optimized for extended,
hard-spectrum sources is sensitive to sources with an integral flux of $6\%$ of
the Crab Nebula above 200 GeV.


Based on analysis of both simulated survey grids and four-telescope Crab data
with one telescope removed from consideration, we estimate that using the
three-telescope array configuration causes at most a $10\%$ loss in
sensitivity, and affects about one third of the total survey region.  The
impact of zenith angle on the survey sensitivity is more variable; based on
current simulations and Crab data, we do not expect the survey sensitivity to
change significantly for regions where the average zenith angle is between
$20^{\circ}$ and $30^{\circ}$, but we do expect a significant drop in
sensitivity for the roughly $12.5\%$ of the survey where the average zenith
angle is greater than $30^{\circ}$.

The interaction of the effects discussed above, as well as the
impact of other potential systematic effects on sensitivity (such as
the the variation in azimuth angle of observations over the survey
region) are still under investigation, but a reasonable estimate
still places the point-source sensitivity at better than $4\%$ of
the Crab Nebula flux over most of the survey region.

\section{Summary}

VERITAS has completed a 140-hour survey of a 5 by 15 degree portion of the
Cygnus region of the Galactic Plane.  Simulation studies indicate that the
survey's average VHE flux sensitivity is better than $4\%$ of the Crab Nebula
at energies above 200 GeV for point sources, and better than $8\%$ of the Crab
Nebula at energies above 200 GeV for extended sources.  Further follow-up
observations have been scheduled for Spring 2009 on the basis of the
preliminary survey analysis, and the final survey maps will be released after
these observations have been completed.

\section{Acknowledgments}

This research is supported by grants from the U.S. Department of Energy, the US
National Science Foundation, and the Smithsonian Institution, by NSERC in
Canada, by Science Foundation Ireland, and by STFC in the UK.  We acknowledge
the excellent work of the technical support staff at the FLWO and the
collaboration institutions in the construction and operation of the instrument.

\end{document}